\documentclass{aimc2026}

\usepackage[utf8]{inputenc} 
\usepackage[T1]{fontenc}    
\usepackage{hyperref}       
\hypersetup{
  colorlinks = true,  
  urlcolor   = blue,  
  linkcolor  = black, 
  citecolor  = black  
}
\usepackage{url}            
\usepackage{booktabs}       
\usepackage{amsfonts}       
\usepackage{nicefrac}       
\usepackage{microtype}      
\usepackage{graphicx}       
\usepackage{amsmath} 
\usepackage{cleveref}       

\usepackage{mdframed}

\newenvironment{affordancenote}{%
  \medskip
  \begin{mdframed}[
    linewidth=0.4pt,
    linecolor=black,
    innerleftmargin=6pt,
    innerrightmargin=6pt,
    innertopmargin=4pt,
    innerbottommargin=4pt,
    skipabove=\medskipamount,
    skipbelow=\medskipamount
  ]
  \small\textit{Affordance note.}\space\ignorespaces
}{\end{mdframed}}

\usepackage{etoolbox}
\newtoggle{anonymous}

\title{Architecture and Affordances of PLAUD: Performative Latents and Unsupervised DDSP}

\iftoggle{anonymous}{%
  \author{%
    Anonymous Author(s)\thanks{Authors and affiliations withheld for double-blind review.}\\ 
    Affiliation withheld for review\\
    \texttt{anonymous@example.com}
    }
}{%
  \author{%
    Błażej Kotowski\thanks{Website of the author: \url{https://blazejkotowski.com}} \\
    Music Technology Group\\
    Universitat Pompeu Fabra\\
    Barcelona \\
    \texttt{blazej.kotowski@upf.edu} \\
    \And
    Frederic Font \\
    Music Technology Group\\
    Universitat Pompeu Fabra\\
    Barcelona \\
    \texttt{frederic.font@upf.edu} \\
  }
}

\begin{document}

\twocolumn[%
  \maketitle
  \begin{abstract}
    PLAUD (Performative Latents and Unsupervised DDSP) is a neural synthesizer and Max for Live instrument for live electronic music, built on NoiseBandNet and trained on small personal sound corpora. We present its architecture, combining a variational DDSP synthesis model, latent smoothing, multi-scale spectral and adversarial losses, and an optional transformer prior, alongside a set of bending operations that intervene directly in the synthesis chain: component limiting, waveshaping, and prior feedback. The Max for Live interface exposes control generation, trajectory sampling, and modulation as primary modes of interaction. Throughout, we thread an affordance analysis arguing that the system's performative character follows from architectural decisions rather than being designed on top of them. The paper contributes both a technical account of the system and a situated affordance analysis of its role in live electronic music performance.
  \end{abstract}%
]

\aimcnotice 

\section{Introduction}
PLAUD — standing for Performative Latents And Unsupervised DDSP — is a real-time generative audio system, which, on the backend, consists of two loosely coupled neural networks trained on shared data: the DDSP-based synthesis engine responsible for the sonic vocabulary, and the Transformer-based prior module, providing its temporal evolution. On the frontend, the interaction layer is implemented through the Max for Live instrument.

PLAUD was developed with electronic music performance in mind and tested on stage throughout the engineering process. Design decisions were taken through first-person experimentation with the system by the first author, and are motivated by an interest in material engagement with machine learning artefacts. AI music generation is now widely available in various forms, from generative platforms offering mostly prompt-based audio generation, to small, DAW-compatible instruments \citep{caillonRAVEVariationalAutoencoder2021, engelDDSPDifferentiableDigital2020, mitcheltreeNeutoneSDK2025}. Such systems are often developed with imitation in mind: imitating qualities of music present in data, or the timbral characteristics of a target acoustic instrument, with control schemes based on well-established semantic and compositional patterns. Creative experimentation, by contrast, points to qualities specific to ML techniques, often contingent on the use of specific architectures, which seem to hold potential for direct engagement in the context of music performance. Previous work exploring these facets of ML engineering in the music domain includes experimentation with latent spaces \citep{tatar_sound_2023} and our own work on network bending \citep{kotowski_2026_bending}, among others. In this article, we subscribe to such approaches, exploring what more direct forms of material engagement with ML might offer in terms of musical expression.

This paper details PLAUD's technical architecture and reflects on the engineering motivations behind it, threading an affordance analysis across the levels of the model. We follow Davis' account of affordances \citep{davisTheorizingAffordances2016, davisAffordancesForML2023}, which complicates the binary understanding of artefacts either possessing or not possessing specific qualities, proposing instead that "artifacts \emph{request}, \emph{demand}, \emph{allow}, \emph{encourage}, \emph{discourage}, and \emph{refuse}" specific interactions. Affordances so understood are situated, in line with Gibson's formulation that they "have to be measured relative to the animal" \citep[p.~127]{Gibson1979}. Here the animal is singular and specific: throughout the paper, affordances are measured relative to the first author, an engineer and performer of experimental electronic music who built the system and is, to date, its only performer. The remarks appear as set-apart \emph{affordance notes}, threading the analysis alongside the technical exposition.

\section{Related Work}
This work builds on both technical research in neural audio synthesis and analytical work on ML materiality and affordances. This section reviews these areas in turn.

\subsection{Neural Audio Synthesis}
Differentiable Digital Signal Processing (DDSP) embedded classical signal processing components within neural network training loops, producing synthesis models whose internal structure remains interpretable and open to intervention \citep{engelDDSPDifferentiableDigital2020}. The paradigm has since expanded into a diverse landscape of instrument-oriented, sound-design, and vocoder applications, surveyed comprehensively by \citet{hayesReviewDifferentiableDigital2023}. NoiseBandNet \citep{barahona-riosNoiseBandNetControllableTimeVarying2023} departs from DDSP's harmonic assumptions by substituting oscillators with a filterbank of narrow noise bands, making the architecture suitable for noisy, textural, and unpitched material while retaining the interpretability of the synthesis chain. In parallel, variational approaches to audio representation have shown that structured latent spaces can organise timbral variation along perceptually meaningful axes \citep{esling2018generative}. On the training side, adversarial losses such as MelGAN's multi-scale discriminator \citep{kumarMelGANGenerativeAdversarial2019} have proven effective at recovering spectral detail that averaging-prone reconstruction objectives suppress.

RAVE \citep{caillonRAVEVariationalAutoencoder2021} represents the most widely adopted instrument-oriented neural synthesizer to date, combining a variational latent space with an optional autoregressive prior and real-time inference through the nn$\sim$ external. Its streaming architecture and small training footprint have made it a common starting point for instrument builders working with neural audio. The present work shares deployment infrastructure with RAVE but departs from it at the synthesis level, operating on a DDSP chain rather than a learned waveform decoder.

\subsection{Latent Space Navigation and Active Divergence}
A growing body of work investigates how latent audio spaces can be navigated and performed in practice. \citet{tatar_sound_2023} propose traversal strategies for sound design contexts that foreground artistic agency over technical optimisation. \citet{tahiroglu_latent_2024} examine GAN latent spaces as platforms for sonic creativity, arguing that their relationship to real-time performance remains problematic. \citet{privato2024stacco}'s Stacco investigates how different models afford distinct modes of gestural control through a dedicated physical interface, while \citet{zheng2025exploring} contribute empirical observations on how performers develop intuitions for abstract parameter spaces. Across these projects a common theme emerges: the structure and mode of access to the latent space determine what the resulting instrument makes possible. Much of the analysis in the present paper is motivated by this question: \emph{how do affordances propagate from architectural decisions through interface design into performance practice?}

A complementary strand engages directly with the breakdown and misuse of generative models. \citet{broadActiveDivergenceGenerative2021} formalise active divergence as a framework for deliberately pushing generative models beyond their training distributions, identifying a taxonomy of strategies including data-level, network-level, and training-level interventions. The framework provides useful vocabulary for practices that treat out-of-distribution behaviour not as failure but as compositional material — a perspective with roots in the aesthetics of failure articulated by \citet{casconeAestheticsFailurePostDigital2000}. In the audio domain, network bending — real-time modification of neural network parameters during sound generation — has been explored as a form of creative misuse grounded in the ethos of circuit bending, treating instability and breakdown as sites of musical invention \citep{kotowski_2026_bending}. A related strategy can be applied to autoregressive generation: feeding a prior's own output back into its context produces controlled departures from learned temporal structure  \iftoggle{anonymous}{\citep{anonymous_brokenforecasts_2025}}{\citep{kotowskiBroken2025}}. These approaches share a commitment to engaging with the specific material behaviours of neural systems rather than abstracting them away behind conventional control paradigms.

\subsection{Materiality and Affordances}
The affordance analysis threaded through this paper draws primarily on \citet{davisTheorizingAffordances2016}'s framework, which moves beyond binary accounts of what artefacts do or do not afford, proposing instead that artefacts \emph{request}, \emph{demand}, \emph{allow}, \emph{encourage}, \emph{discourage}, and \emph{refuse} specific interactions depending on the user and context. \citet{davisAffordancesForML2023} extends this framework to machine learning systems, treating ML applications as objects of design whose features invite situated affordance analysis. The present work applies this lens to ask how the architectural and training-time decisions shape a neural synthesiser's affordances in performance. This situated understanding aligns with Gibson's ecological formulation, in which affordances are always relative to the organism that encounters them \citep{Gibson1979}.

Several perspectives inform the approach taken here. Technological mediation theory grounds how artefacts actively shape the practices they participate in \citep{verbeek2005things}, a stance extended in postphenomenological studies of computational systems and in analyses of how specific architectural decisions yield distinct situated affordances across contexts \citep{gerlekMaterialityMachinicEmbodiment2025,kotowskiExploringSituatedStabilities2025}. \citet{benjaminMachineLearningUncertainty2021} reframe ML uncertainty as a design material rather than an obstacle. In sound, \citet{fellStructureSynthesis2022} argues from direct practice that tool-making and creative work are inseparable, and \citet{scurtoPrototypingMachineLearning2021} that small-data regimes and first-person experimentation are legitimate modes of inquiry into ML — the methodological stance adopted here.

\section {System Architecture}
PLAUD's architecture consists of two loosely coupled components: a DDSP-based neural synthesizer with a variational latent space, and an optional autoregressive prior. Together they define the instrument's sonic vocabulary and temporal tendencies.

\subsection {NoiseBandNet-Based Synthesis Model}\label{subsec:architecture_synthesis}
The synthesis model is based on NoiseBandNet \citep{barahona-riosNoiseBandNetControllableTimeVarying2023}, a DDSP architecture replacing harmonic oscillators with a filterbank of narrow noise components. The output is a weighted sum of $M$ pre-rendered noise bands:
$$
y(t) = \sum_{m=1}^{M} a_m(t) \cdot n_m(t)
$$
where $n_m(t)$ is the $m$-th band and $a_m(t)$ its predicted amplitude. The bands are pre-computed by filtering white noise through $M$ adjacent FIR filters tiling $[0, F_s/2]$ and stored as loopable wavetables. This amounts to additive synthesis with noise bands as base functions, preserving interpretability while avoiding inductive bias toward harmonicity. 

\begin{affordancenote}
    The transparency of the synthesis chain — fixed noise bands, predicted amplitudes — \textit{allows} direct intervention in the signal path. This interpretability \textit{encourages} creative manipulation that would be opaque in less structured approaches, such as neural codec-based synthesis. Several such interventions are described in \cref{subsec:interaction_bending}.
\end{affordancenote}

NoiseBandNet was conceived for sound effects, where pitch-based assumptions do not hold. This makes it a great candidate for a universal audio synthesis architecture. The model predicts at an internal rate $F_s/W$ and upsamples through linear interpolation. $W$ defines temporal resolution, while $M$ affects reconstruction quality and model size. In its original formulation, NoiseBandNet uses pre-engineered features (loudness, spectral centroid) rather than a learned latent space — departing from DDSP's encoder-derived $z$. This suits extremely compact datasets ($\sim$4 to $\sim$95 seconds in the original paper), where sonic variance is adequately captured by a few acoustically grounded features.

\begin{affordancenote}
This small-data regime \textit{encouraged} engagement with compact, hand-curated collections, steering early development toward treating corpus curation as a compositional act.
\end{affordancenote}

Experiments with fitting more complex material (i.e. ~30 minutes of drone music composed by the first author) revealed that loudness and centroid constrain the model to dominant timbral states: overlapping regions in the feature space force the network to average across perceptually distinct sounds. We substituted these inputs with a 4-dimensional variational latent space, letting the encoder discover parameters describing the data distribution on its own. Although in the original DDSP formulation, the encoder is preserved for style-transfer, we use it exclusively to learn the latent space during the training; at the inference time, only the decoder is used.

\begin{figure*}[ht]
  \centering
  \includegraphics[width=\linewidth]{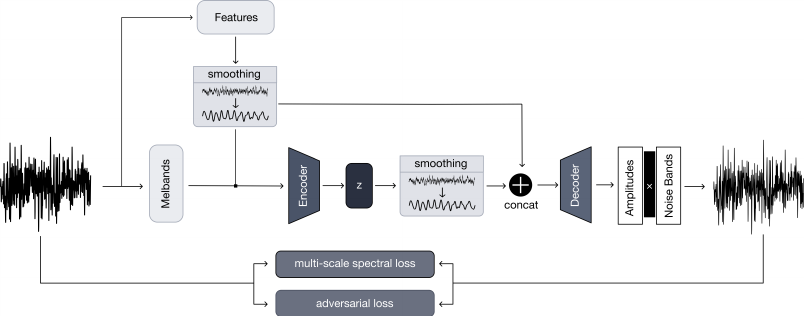}
  \caption{
    PLAUD synthesis model architecture. During training, mel-band features feed the encoder (producing latent $z$), while raw audio feed a parallel feature extraction path; both are smoothed before concatenation and decoding into per-band amplitudes, which are multiplied with pre-rendered noise bands. Training combines multi-scale spectral and adversarial losses. In inference, only the decoder is used. The prior network is not shown.
  }
  \label{fig:plaud_architecture}
\end{figure*}

\begin{affordancenote}
Latent navigation \textit{allows} a more exploratory control approach, shifting the performer's relationship from interpretive adjustment of known quantities toward discovery within an abstract space whose meaning is learned through practice.
\end{affordancenote}

Live performance later motivated reintroducing loudness and centroid alongside the latents, now reduced to two dimensions. The resulting hybrid control space provides minimal perceptual anchoring while retaining the exploratory character of the latent space.

\subsection{Latent Space Regularisation}\label{subsec:architecture_vae}
The latent space in the original DDSP architecture is unregularised, allowing the encoder to organise representations arbitrarily. In practice, this produces latent topologies with narrow, concentrated clusters separated by unpopulated gaps. In such a topology, spectrally similar timbres may end up distant in Euclidean terms, and small parameter changes can cross empty regions to produce disproportionate sonic jumps. We address this by training the encoder as a variational autoencoder \citep{kingmaIntroVAE2019}, adding a KL divergence term to the loss, weighted by a factor $\beta$ \citep{burgess2018understanding}:

\begin{equation}
\mathcal{L} = \mathcal{L}_{\text{reconstruction}} + \beta \cdot D_{\text{KL}}(q(z|x) \| p(z))
\end{equation}

\begin{figure}[ht]
  \centering
  \includegraphics[width=\linewidth]{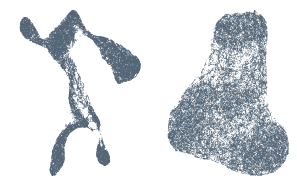}
  \caption{
    Latent space projections from models trained on the same corpus with $\beta = 10^{-3}$ (left) and $\beta = 10^{-2}$ (right). Stronger regularisation yields a denser, more uniformly populated manifold, reducing the unpopulated gaps that make continuous navigation unpredictable.
  }
  \label{fig:beta_effect}
\end{figure}

Higher $\beta$ values push the encoder toward a denser, more uniformly populated latent manifold (\cref{fig:beta_effect}). The effect is directly relevant to performance: a space with fewer gaps means that continuous movement in any direction is more likely to produce continuous timbral change, better satisfying established criteria for digital musical instrument design where parameter changes should yield proportional sonic responses \citep{jorda2004digital_pt2}. Conversely, lower $\beta$ values preserve more of the data's intrinsic structure at the cost of navigability --- the resulting space is more expressive in principle, but harder to traverse smoothly.

\begin{affordancenote}
    The regularised, continuous latent space \textit{encourages} exploratory navigation by making the space more uniformly responsive to control input. A space with fewer gaps means that small changes in control are more likely to produce correspondingly small changes in the output, a property \citet{jorda2004digital_pt2} identifies as necessary for the development of expressive control. Conversely, lower $\beta$ values \textit{discourage} casual traversal by concentrating representational density in specific regions, demanding more precise, deliberate control from the performer.
\end{affordancenote}

In training, the $\beta$ factor is increased from 0 to its target value across epochs 100--200 in a warmup schedule, avoiding premature posterior collapse that would flatten the latent space before the decoder has learned useful reconstructions.

\subsection{Adversarial Training}
Multi-scale spectral (MSS) losses tend to produce over-smoothed synthesis parameters, suppressing fine spectral detail — an effect demonstrated in the DDSP context by \citet{wuetalMidiDDSP2022}. We observed the same training PLAUD on noisy, detail-rich textures. Additionally, spectral losses fall short in optimising for pitch correctly \citep{turianSorryForYouLoss2020}. Adding a MelGAN-style adversarial feature loss \citep{kumarMelGANGenerativeAdversarial2019} helps recovering spectral detail, especially in higher parts of the spectrum. The discriminator consists of three sub-discriminators, each operating at a progressively 4$\times$ downsampled scale, with three layers each, producing a multi-scale feature matching loss. The DDSP model serves as the generator.

\begin{figure}[ht]
  \centering
  \includegraphics[width=\linewidth]{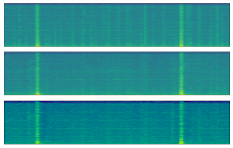}
  \caption{
    Log-magnitude spectrograms of training audio (top), reconstruction without adversarial training (middle), and reconstruction with adversarial training (bottom). The adversarially trained model recovers finer spectral detail, visible as sharper horizontal structures in the upper frequency range and increased contrast in transient regions, compared to the smoother, more diffuse reconstruction produced by MSS loss alone.
  }
  \label{fig:adversarial_effect}
\end{figure}

Notably, adversarial training can worsen the MSS reconstruction metric while improving perceptual quality — the discriminator pushes toward spectral detail that averaging-prone reconstruction losses actively suppress (see \cref{fig:adversarial_effect}). To avoid instability, adversarial training activates after 200 epochs of MSS+KLD training, functioning as perceptual fine-tuning rather than a primary objective.

\begin{affordancenote}
    Adversarial training allows inclusion of more texturally rich, noisy training material that MSS-only regimes discourage, broadening the range of sound corpora the system can meaningfully learn from.
\end{affordancenote}

\subsection{Latent Smoothing and Short-Term Temporal Context}\label{subsec:architecture_smoothing}
At PLAUD's default resampling rate of $W=32$ at 44.1kHz, the control signal operates at ~1378 Hz — fine-grained enough that the encoder produces highly dynamic latent trajectories. Reproducing such sequences manually is unfeasible, making it difficult to reproduce the temporal structures embedded in the data through direct latent navigation.

Inspired by the smoothing strategy in Sketch2Sound \citep{sketch2sound2025}, we smooth the latent trajectories during training using a moving average filter:

$$\tilde{z}_l(t) = \frac{1}{K} \sum_{k=0}^{K-1} z_l(t+k), \quad l = 1, \ldots, L$$

where $L$ is the number of latent dimensions and $K$ the kernel size. This removes high-frequency content from the trajectories, encouraging the GRU of the decoder to internalise more of the short-term temporal structure. Where in standard DDSP the GRU encodes minimal context, smoothing pushes it to preserve longer continuity — the same latent point will sound differently depending on the trajectory leading to it. This frees the optional prior network (\cref{subsec:architecture_prior}) to operate at a higher temporal level (\cref{subsec:architecture_tradeoff}).

\begin{figure}[ht]
  \centering
  \includegraphics[width=\linewidth]{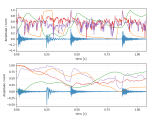}
  \caption{
    Control trajectories over a ~1 second excerpt, without smoothing (top) and with smoothing at kernel size $K=257$ (bottom). Each coloured line represents one control dimension; the blue waveform shows the corresponding audio. Without smoothing, the trajectories are highly dynamic and tightly coupled to the audio's micro-temporal structure. After smoothing, the same dimensions trace slower, more gestural contours — delegating short-term continuity to the GRU and making the latent space navigable through simple modulation.
  }
  \label{fig:smoothing_latents}
\end{figure}

\begin{affordancenote}
Smoothing \textit{allows} synthesis of temporally complex structures from relatively simple latent modulation (see \cref{fig:smoothing_latents}). The GRU enacts movement once a position in the latent space is established. This inspired the modulation-based control mode described in \cref{subsubsec:interaction_modulation}. A related idea of leveraging modulation patterns within DDSP has been explored independently by \citet{mitcheltreeetalModulation2025}.
\end{affordancenote}

\subsection{Transformer-Based Autoregressive Prior}\label{subsec:architecture_prior}
\begin{figure*}[ht]
  \centering
  \includegraphics[width=\linewidth]{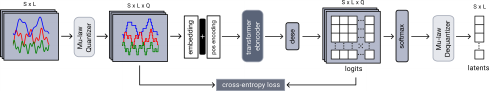}
  \caption{
    Prior network architecture. Continuous control signals are mu-law quantized into discrete tokens, embedded with positional encoding, and processed by an encoder-only transformer. A dense layer produces logits over the quantization classes, trained with cross-entropy loss; at inference, softmax sampling and mu-law dequantization recover continuous control trajectories.
  }
  \label{fig:prior_architecture}
\end{figure*}

The prior is an optional component, meaning that the synthesizer – controlled directly through parameter modulation – is fully functional without it. In training, the prior learns temporal structure from the corpus: rhythmic patterns, envelopes, and longer-range sequential dependencies that the GRU-based decoder does not capture.

The architecture is an encoder-only transformer operating autoregressively over the synthesizer's discretised control space. Following a similar approach to the RAVE prior \citep{caillonRAVEVariationalAutoencoder2021}, the continuous control signals are quantized into $Q$ classes per latent dimension through mu-law companding:

$$c_q = \text{round}\left(\frac{Q-1}{2} \cdot \frac{\ln(1 + \mu |z|)}{\ln(1 + \mu)} \cdot \text{sgn}(z) + \frac{Q-1}{2}\right)$$

The algorithm nonlinearity allocates finer resolution near zero, matching the concentration of values around the VAE's approximately normal distribution. We train with $Q=32$ classes, which provides sufficient resolution to reconstruct meaningful control trajectories while keeping the vocabulary compact. The transformer receives quantized codes as token embeddings with positional encoding and predicts the next token per dimension. It is trained via cross-entropy loss. At inference, logits pass through softmax and are dequantized back to continuous values. Since models share the same control parameter space, any prior can drive any synthesizer. Pairing synthesisers with priors trained on distinct corpora produces a loosely interpreted form of style transfer: the temporal structure of one dataset — its phrasing, rates of change — is imposed onto the timbral vocabulary of another, or the other way around.

\begin{affordancenote}
Combining priors and synthesizers from distinct corpora \textit{allows} cross-corpus composition. The softmax temperature \text{allows} balancing between closer reproduction of learned patterns and increasingly stochastic generation.
\end{affordancenote}

\begin{affordancenote}
The autoregressive nature of the prior — generating one token at a time conditioned on its own recent output — allows for feedback interventions where generated sequences are folded back into the transformer's context. The mechanism is detailed in \iftoggle{anonymous}{\citet{anonymous_brokenforecasts_2025}}{\citet{kotowskiBroken2025}}, where destabilising the prior's self-conditioning produced patterns that depart from the learned temporal structure in controlled ways. The technique was subsequently integrated into PLAUD's interface as a performance control (\cref{subsec:interaction_bending}).
\end{affordancenote}

\subsection{Resampling Rate Trade-off}\label{subsec:architecture_tradeoff}
The resampling rate $W$ (the number of audio samples per control frame) determines how often the model predicts, and consequently, how much temporal structure must be delegated to the prior versus resolved directly by the synthesizer. At audio rate $F_s$, the internal control rate is $f_{int} = F_s / W$, meaning each predicted amplitude frame spans $\Delta t = W / F_s$ seconds. For a prior with a fixed sequence length of $L$ tokens, the duration of audio the prior can condition on equals:

$$T_{RF} = \frac{L \cdot W}{F_s}$$

\begin{figure}[ht]
  \centering
  \includegraphics[width=\linewidth]{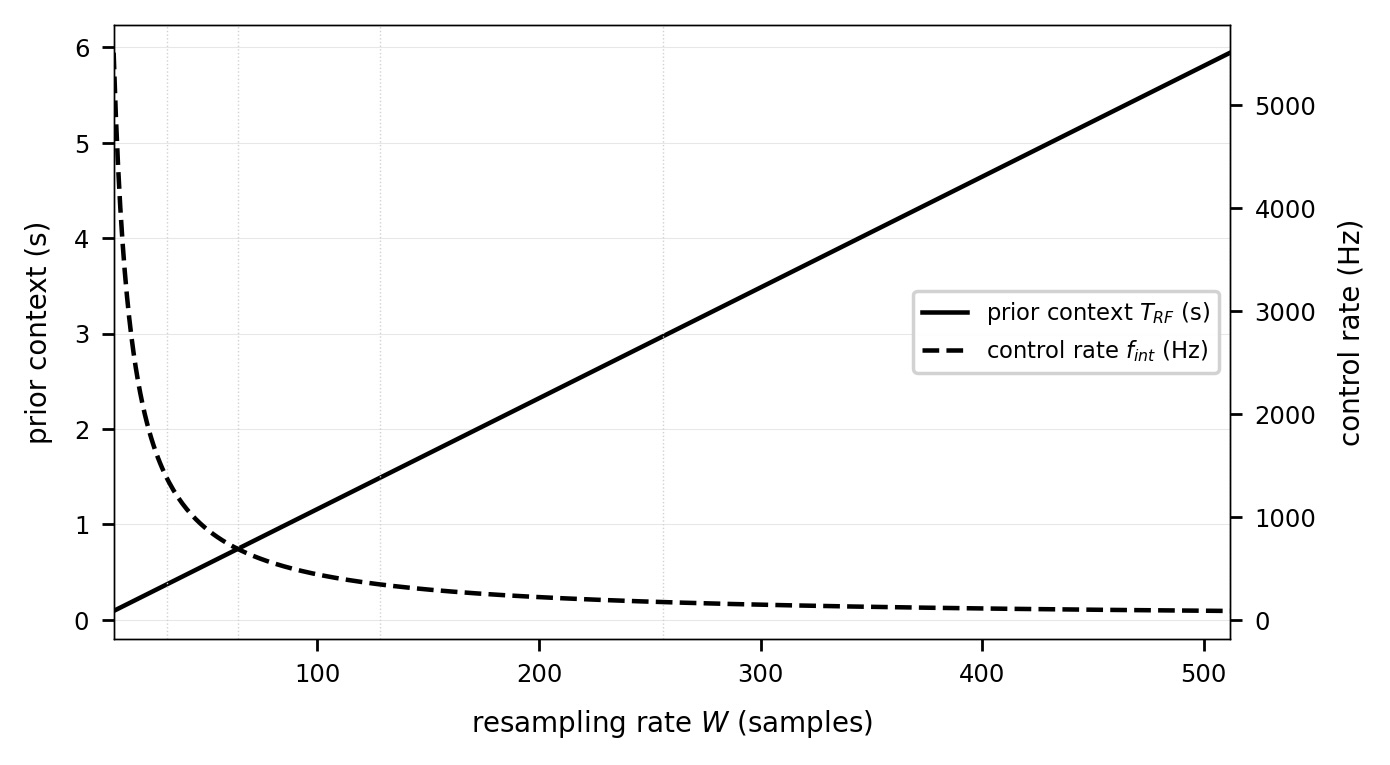}
  \caption{
   Prior temporal context $T_{RF}$ (solid) and internal control rate $f_{int}$ (dashed) as functions of resampling rate $W$, for a fixed prior sequence length $L = 512$ at $F_s = 44100$ Hz. Their product is constant at $L$.
  }
  \label{fig:resampling_tradeoff}
\end{figure}

The control rate and the prior's temporal context move in opposite directions. Lower $W$ yields a higher $f_{int}$, resulting in sharper transients, and less envelope smearing. At the same time it produces more tokens per unit time, making the prior's sequence modelling task harder and more compute-intensive. Higher $W$ extends $T_{RF}$, but coarsens the synthesis envelope, blurring rapid timbral changes. The two quantities are bound by a fixed product: $T_{RF} \cdot f_{int} = L$, and cannot be independently optimised (see \cref{fig:resampling_tradeoff}).

There is no universally optimal setting. The choice is tuned to the corpus and the intended use. Texturally dense material demands lower $W$, while slowly evolving textures tolerate higher values that grant the prior a wider temporal view.

\section{Interface and Interaction}
The Max for Live instrument exposes PLAUD's latent space, autoregressive prior, and synthesis chain through a set of controls and interaction modes, detailed in the following subsections\footnote{A companion page with video demonstrations of the interactions discussed in this section is available at \url{https://plaudaimc2026.github.io/}.}. The device is designed to engage the material affordances of the underlying networks described in the previous sections, and can in fact be seen as emerging from these affordances as much as from the performance intentions of the first author.

\subsection{The Max for Live Instrument}
\begin{figure*}[ht]
  \centering
  \includegraphics[width=\linewidth]{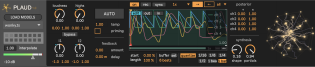}
  \caption{
    The PLAUD Max for Live interface. From left: model loading, presets and interpolation controls; loudness, highs, and latent dimension control knobs; autoregressive prior controls, latent trajectory and sampling; posterior trajectory modifiers and synthesis bending parameters; bending visualizer.
  }
  \label{fig:max4live_interface}
\end{figure*}

PLAUD is implemented as a Max for Live device, built around the nn$\sim$ external\footnote{See \url{https://github.com/acids-ircam/nn_tilde}.}, originally developed for deploying IRCAM's RAVE models to Pure Data and Max. The synthesizer and prior networks are bundled and exported into an nn$\sim$-compatible format.

\begin{affordancenote}
    The reliance on nn$\sim$ \textit{encourages} adoption among users who have already experimented with RAVE-based instruments, as no additional externals are needed. The choice of Max for Live \textit{allows} seamless extension through Live's modulation routing, automation, and effect chaining, and \textit{discourages} viewing PLAUD as a closed, self-contained system.
\end{affordancenote}

The interface is organised in functional blocks from left to right (See \cref{fig:max4live_interface}). The leftmost section handles model and preset management: up to sixteen parameter snapshots can be saved and interpolated between. Below, a volume control sets the output level and a meter provides visual monitoring. Four control knobs are mapped to the synthesis model's input parameters: loudness, spectral centroid, and two latent variables (\cref{subsec:architecture_synthesis}). Default ranges are $[0, 1]$ for audio features and $[-1, 1]$ for latents, reflecting training data distributions. These ranges are adjustable, allowing the performer to push the model beyond its training manifold into regions where it must interpret unseen control signals. A bypass toggle disables the knobs temporarily.

\begin{affordancenote}
    Adjustable control ranges \textit{allow} the performer to probe the latent space beyond its training distribution and \textit{encourage} experimentation with out-of-distribution behaviour.
\end{affordancenote}
    
The prior control section features an "auto" toggle for engaging the autoregressive prior, temperature and priming controls (\cref{subsubsec:interaction_prior}), and feedback parameters (\cref{subsec:interaction_bending}). Adjacent is the trajectory sampling module (\cref{subsec:interaction_trajectory_sampling}). The posterior modifiers block allows offsetting and scaling the control streams — whether from the looper or the knobs — alongside two bending controls for waveshaping the synthesis basis functions and adjusting the number of active noise bands (\cref{subsec:interaction_bending}). Finally, an abstract visualisation renders a sphere of spokes encoding waveshaping value, partial count, and playback speed.

\subsection{Two Sources of Control Signal}
The synthesizer's input parameters can be driven by two complementary sources: external modulation and control generation by the prior.
\subsubsection{Modulation}\label{subsubsec:interaction_modulation}
As a Max for Live device, PLAUD's parameters are available to any modulation source within Ableton Live: LFOs, sequencers, envelopes, or manual control. These sources produce control trajectories the model never encountered during training.

\begin{affordancenote}
Modulation is a familiar gesture in synthesizer practice, making this interaction mode immediately accessible to electronic musicians. It allows integration of PLAUD into existing modulation workflows.
\end{affordancenote}

\subsubsection{Control Generation by the Prior}\label{subsubsec:interaction_prior}
When the autoregressive prior is engaged, it generates control trajectories in real time, automating the control parameters according to temporal patterns learned from the training data: learned expressive patterns and longer-term temporal dynamics that would be difficult to reproduce manually.
The prior can be primed by manual modulation: a short segment of trajectory is fed as context, after which the network continues from where the input left off. This allows the performer to seed generation with a specific starting condition and let the prior develop it further. Temperature controls the softmax distribution over the quantised control vocabulary, controlling predictability of generation. A feedback parameter routes the prior's own output back into its context, described further in \cref{subsec:interaction_bending}.

\subsection{Trajectory Sampling}\label{subsec:interaction_trajectory_sampling}
Trajectory sampling mirrors audio sampling techniques such as looping, slicing, and transposition, but operates in the control space rather than on audio directly. The module records control trajectories into a buffer, capturing the sum of knob positions and prior generation (if enabled). Playback supports forward and reverse directions with adjustable speed and buffer length. Start and end points define a loop region within the buffer. Two heads — playback and recording — can either move in sync or independently: the record head loops around the entire buffer while playback cycles between the selected start and end points. Operations can be quantised to a grid at selectable resolutions. A real-time preview displays the generated parameter modulation across up to four channels, and the functionality can be applied to a single selected channel or all at once.

Although the interaction vocabulary resembles audio sampling, the nature of neural synthesis augments the generation. Because the smoothed decoder makes the synthesised output depend on the trajectory leading to a given latent point and not only on the point itself (\cref{subsec:architecture_smoothing}), playback speed and direction shape the emerging temporal structures rather than simply compressing or stretching them. The posterior modifiers shift timbral properties of the synthesised output or relocate it within the training data manifold. Quantisation locks operations to the session grid, integrating trajectory sampling into the broader project context.

\begin{affordancenote}
The familiarity of sampling controls among electronic music producers \textit{encourages} transferring existing intuitions about looping, slicing, and speed adjustment to the control domain. At the same time, the differences arising from neural synthesis \textit{allows} exploration beyond what audio sampling affords.
\end{affordancenote}

\subsection{Bending Operations}\label{subsec:interaction_bending}
Three operations intervene directly in the synthesis chain and prior generation, collectively referred to as bending operations. The term draws from circuit and network bending \citep{kotowski_2026_bending}.\footnote{Whether these interventions constitute network bending in a strict sense is debatable, as they operate on the synthesis chain rather than on weights or activations. However, we believe the term fits broader interpretations and contributes to situating these operations within a lineage of creative misuse.}

\emph{Component limiting} reduces the number of active noise bands used for synthesis. At each frame, the bands with the highest predicted amplitudes are retained and the rest silenced. This produces abrupt spectral discontinuities that occasionally result in clicks, an artefact that proved musically interesting in practice and was therefore kept. 

\emph{Waveshaping} substitutes noise bands with frequency-matched sinusoids and progressively shapes them toward square waves, producing a continuum from noise through sine to square. Both operations exploit the interpretability of the DDSP synthesis chain.

\emph{Prior feedback} routes the generated latent codes back into the prior's own predictions, with adjustable amount and delay, destabilising generation implementing the mechanism detailed in \iftoggle{anonymous}{\citet{anonymous_brokenforecasts_2025}}{\citet{kotowskiBroken2025}}.

\begin{affordancenote}
    Bending operations \textit{allow} moving sound beyond the training data space in a controlled, musically legible way. They \textit{encourage} direct engagement with the materiality of the network.
\end{affordancenote}

\section{Performance Reflection}\label{sec:performance_reflection}
\begin{figure}[ht]
  \centering
  \includegraphics[width=\linewidth]{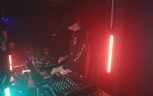}
  \caption{
    The first author performing with PLAUD in a live club setting. The setup includes a laptop running Ableton Live with the PLAUD Max for Live device, alongside external Push controller.
  }
  \label{fig:plaud_performance}
\end{figure}

PLAUD has been used in a range of live electronic music contexts, from open improvisation to more structured sets with prepared presets and loose scores, across four public performances by the first author, who remains its only performer to date. This section reflects on what the system makes possible in performance, and what it resists.

Working with self-curated datasets is central to the practice. Quick training times and small data requirements mean that a different set of models, trained on new material, can be prepared for each performance. This produces performance-specific instruments that differ in reactivity, timbral vocabulary, and overall character. No two are the same, which keeps the process surprising and generative as a creative practice.

The bending operations have proven to be expressive performance tools. They allow dramatic shifts: limiting the number of active components to the minimum produces minimal compositions of unstable sine tones and clicks. Waveshaping these minimal signals introduces synthetic precision, and the full spectral detail of the original model can be restored with a single knob movement.

A recurring observation concerns the relationship between reconstruction accuracy and expressivity. Neither the synthesizer nor the prior achieves high-fidelity reproduction of the training data, and in many cases the reconstruction quality would not satisfy conventional evaluation criteria. This did not prove to be an obstacle. For the synthesizer, the lo-fi character of the output became part of the instrument's identity. For the prior, stable long-term generation was never achieved, because of accumulating prediction errors leading to drifts. In a creative context where faithful reproduction is not the goal, this instability can be seen as an aesthetic affordance. The resulting sound has a character that can be valued in performance as a distinct quality of the neural synthesis, much as certain iconic synthesizers are sought out for their sonic feel.

The trajectory sampling module was the most recently introduced element, and partly serves to integrate PLAUD's generation into more conventional grid-based setups. In practice it has worked well, providing a way to impose structure over the prior's more chaotic tendencies. One effective technique is to load a model trained on rhythmic data, sample its prior generation into the buffer, then switch to a model trained on different material (e.g. metal impacts), and let the pre-sampled trajectories steer the new synthesis. This produces a form of style transfer performed live, combining the temporal structure of one corpus with the timbral vocabulary of another.

One significant limitation is computational performance. Without the prior, up to three model instances can run simultaneously on a MacBook M1 Pro from 2021 with relative stability, but engaging the prior reduces this to one, leaving limited headroom for additional instruments and effects in a Live session. On several occasions this has required compromises in the performance setup, and in a workshop context some participants' hardware could not run the models at all. Reducing this computational footprint remains the most pressing area for future development.

\section{Conclusion, Limitations and Future Work}
PLAUD demonstrates that the expressive affordances of a neural instrument can emerge directly from its architecture: from the constraints of small data training, the structure of the variational latent space, the interpretability of the DDSP synthesis chain, and the temporal tendencies encoded by the prior.

The paper presented a system combining a VAE-regularised NoiseBandNet synthesis model, latent smoothing, adversarial training, and an optional transformer prior, alongside bending operations that intervene directly in the synthesis chain. The Max for Live instrument exposes these components through trajectory sampling, modulation-based control, and prior's automatic steering mode. Throughout, we have argued that the instrument's performative character follows from its architectural decisions rather than being designed on top of them.

The system's audience and aesthetic commitments are both narrow by design. Building on nn$\sim$ situates PLAUD among setups already prepared for RAVE, and packaging it as a Max for Live device rather than a standalone application addresses producers and performers rather than researchers. Sonically, the model was developed for noisy, textural, and unpitched material, and its lo-fi reconstruction, spectral discontinuities, and drifting prior are treated as instrument character rather than as defects to be engineered away, which suits some practices considerably better than others.

Several directions remain open. Computational cost is the most pressing limitation, as discussed in \Cref{sec:performance_reflection}. On the synthesizer side, the autoregressive GRU is likely a bottleneck. Replacing it with causal convolutions could improve throughput while also offering more explicit control over short-term context length, with potentially interesting architectural affordances of its own. On the prior side, the resampling rate trade-off (\cref{subsec:architecture_tradeoff}) makes it unfeasible to learn structure at low resampling rates while keeping compute manageable. An intermediary rate reduction block, in the style of neural audio codecs \citep{soundstreamZeghidour2022, descriptDACKumar2023} applied to the latent signal, could decouple the synthesizer's temporal resolution from the prior's context length, enabling finer detail without proportionally increasing the prior's computational load.

Alternative DDSP decoders such as sinusoidal-plus-stochastic modelling would broaden the timbral range. Preliminary experiments have shown some promise, but unsupervised learning of frequencies present in the signal remains a hard problem \citep{turianSorryForYouLoss2020}, and although promising research exists \citep{sinusoidalFreqHayes2023}, more work is needed before such decoders can be integrated into the full pipeline. 

Finally, the bending vocabulary could be expanded with operations such as spectral rolling, alternative modes of component subset selection, and harmonisation.

\section*{Author Declarations}
The authors used Claude for language editing to improve grammar and clarity. All scientific content, interpretations, and conclusions were developed by the authors.


\bibliographystyle{apalike}
\bibliography{references}

\end{document}